# Fully Atomistic Molecular Dynamics Investigation of the Simplest Model of Dry-Draw Fabrication of Carbon Nanotube Fibers

Luís F. V. Thomazini and Alexandre F. Fonseca

*Applied Physics Department, "Gleb Wataghin" Institute of Physics, University of Campinas - UNICAMP, Campinas, São Paulo, CEP 13083-859, Brazil.*

## ABSTRACT

*Macroscopic assemblies of carbon nanotubes (CNTs) are desirable materials because of the excellent CNT properties. Amongst the methods of production of these CNT materials, the dry-draw fabrication where CNT fibers (CNTFs) are directly pulled out from a CNT forest is known to provide good physical properties. Although it is known that vertical alignment of CNT bundles within the CNT forest is important, the mechanisms behind the dry-draw fabrication of CNTFs are still not completely understood. The simplest known dry-draw model consists of CNT bundles laterally interacting by only van der Waals forces (vdWf). Here, by fully atomistic classical molecular dynamics simulations, we show that the simplest dry-draw model does not produce CNTFs. We also show one important condition for a pair of adjacent CNT bundles to connect themselves under vdWf only and discuss why it leads to the failure of the simplest model.*

## INTRODUCTION

Carbon nanotubes (CNTs) [1] are single- or multi-walled hollow cylinders at nanoscale. Their structure can be thought as been formed by planar hexagonal networks of carbon atoms rolled up around a certain axis. Depending on how the hexagonal network is cut and rolled up, each wall of the CNT presents unique values of diameter and chirality [2]. CNTs present excellent properties as high elastic modulus and tensile strength (1 TPa and 100 GPa, respectively [3]), high electrical current density (larger than $10^9$ A/cm$^2$ [4]) and high thermal conductance (about 3500 Wm$^{-1}$K$^{-1}$ [5]). These special physical properties make CNTs suitable for several scientific and commercial applications [2,6]. At nanoscale, we have, for example, the development of a single CNT-based functional radio receiver [7], field-effect transistors [8], drug delivery carriers [9,10], efficient solid refrigerants [11,12], energy harvesters [13,14], etc.

One challenge in Nanotechnology is to produce macroscopic nanostructured materials that possess the most of the special properties of their constituent nanostructures. In the particular case of CNTs, one way to fulfil this challenge is find out ways to assembly lots of CNTs into a nanostructured material. They could be produced by dispersing CNTs within other materials like in composites [6,15], or directly processed to form sheets, yarns and fibers of CNTs [16-19]. One particular way to produce CNT fibers (CNTFs) is through pulling them out of a CNT forest or array. It is called forest spinning (FS) dry-draw method [17,19]. The reader is referred to the Review articles [16-19] with regard to the description and details of this and other methods to produce CNTFs. The purpose of this work is to investigate the simplest mechanism to produce CNTFs through FS method.

First obtained by Jiang, Li and Fan [20] and further improved by Zhang, Atkinson and Baughman [21], the FS method consists of pulling out CNTFs from vertically aligned CNT bundles in CNT forests. Other scientific groups further developed the method [22-25]. Basically, there are three models of the structure of CNTs within a CNT forest and the mechanism of conversion from them to horizontally aligned CNTFs: 1) vertically aligned CNT bundles laterally interacting by van der Waals (vdW) forces only [22,23]; 2) vertically aligned CNT bundles connected by individual CNTs or small bundles [24]; and 3) self-entanglement of vertically aligned CNT bundles [25]. For the sake of simplicity, these models will be called M1, M2 and M3, respectively. Among these models, the simplest one is the M1 whose proposed CNT structure is shown in figure **1**. The study and analysis of models M2 and M3 will be subject of a future publication.

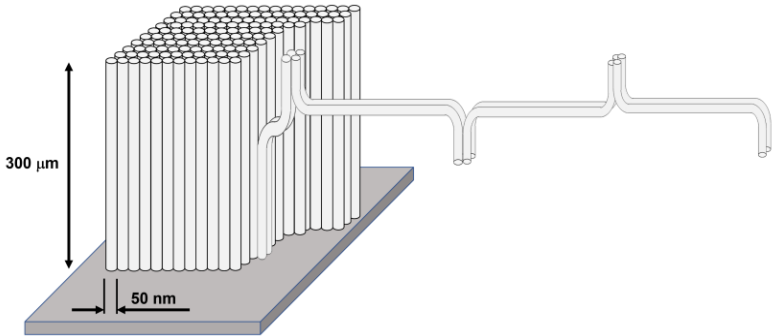

**Figure 1.** Illustration of the M1 model of a vertically aligned CNT forest having one CNTF being pulled out to the right. Cylinders represent bundles of CNTs. This drawing was made based on the figure 2 of Ref. [22]. Typical sizes of a drawable CNT forest are also indicated [21,24].

Figure **1** depicts a vertically aligned CNT forest being pulled out to form a piece of CNTF. It was drawn based on figures 2 of Ref. [22] and 12 of Ref. [23]. This is the simplest model for the mechanism of the FS method because it relies only on lateral vdW interactions between large CNT bundles. Although the M1 looks to resemble the experimental observations of the FS method, it is not clear how vdW forces alone are capable to keep a CNT bundle being pulled out connected to the next one instead of not simply detaching one from each other at their extremities. vdWs forces per contact area are the same on both sides of the vertical CNT bundle. As the details of this process at atomic scale remain unclear, in order to gain insights on these models, we decided to carry out fully atomistic molecular dynamics (MD) simulations using the AIREBO [26] potential as available in the LAMMPS [27] computational package. Based on numerical experiments, we show that the M1 model is not capable to explain the FS process where vertically aligned CNT bundles are continuously converted to horizontally aligned CNTFs. As the issue of the vdW forces on both sides of a CNT bundle is relevant, we also tested an idea of letting a space between some of the vertical bundles at around the extremities of the CNT forest, so as to allow it to be bent towards the pulling out direction.

## THEORY AND SIMULATION DETAILS

Although classical MD allows for simulations of millions of atoms, simulations of a real CNT forest are not practical in view of their enormous size. We, then, chose an approach based on what Machado, Legoas and Galvão [28] used to simulate the FS method. Among different approaches defined by them to represent the CNT forest, we chose that that consists of representing the contact region between adjacent large CNT bundles by two sets of few graphene layers as illustrated in figure **2**. This is justified by the fact that the typical size of the diameter of the CNT bundles in drawable forests is about 50 nm and the CNTs are multiwalled with 9 walls in average per tube [21,24]. Although it is known that large diameter CNTs can easily collapse [29], multi-walled CNTs are expected to be stiff. Our structure model possesses 149904 atoms.

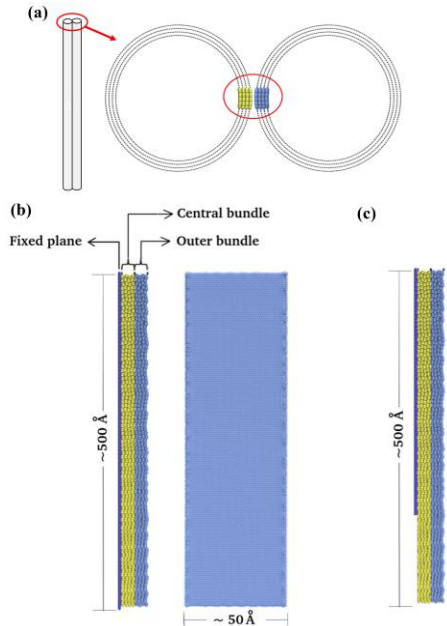

**Figure 2.** (a) Two adjacent CNT bundles and a magnified upper view of the contact between them. Yellow and blue spheres highlight the two regions of contact where the bundles are approximately planar and will be represented by graphene layers in our simulations. (b) Lateral and front views of the atomistic model of two CNT bundles in close contact represented by four graphene layers each. The outer bundle (blue) is the one that will be directly pulled out. Central bundle (yellow) is the one that will be tested regarding being or not pulled out by the outer one. There is an additional fixed graphene plane representing the opposite side of the central bundle in the CNT forest. (c) The same model but with a cut part in the fixed plane to represent an empty space between the central bundle and the rest of the forest.

Classical MD simulations of the structures shown in figure **2** were performed with the AIREBO potential [26] as available in LAMMPS [27]. The structures were first geometry optimized by minimization methods (with force and energy tolerances of $10^{-8}$ eV/Å and $10^{-8}$) followed by equilibrium MD simulations at 300 K for, at least, 100 ps using a Langevin thermostat. Time step and thermostat damping factor were set in 0.5 fs and 1 ps, respectively. No periodic boundary conditions were considered. The FS method is simulated as follows. The outer bundle will be externally pulled out by assigning an initial constant velocity to a set of atoms located at its upper 10 Å. Three velocities values were considered: 0.01, 0.1 and 1.0 Å/ps. The central bundle is the one that will be probed if it will be or not pulled out by the outer one due to only vdW forces between them. One additional graphene plane is included close to the other side of the central bundle to mimic the rest of the CNT forest.

**RESULTS AND DISCUSSION**

As observed by Machado, Legoas and Galvão [28], because of low friction between graphene layers, the outer bundle can detach and slip depending on the pulling out velocity. In our case, for the lowest pulling out velocity, 0.01 Å/ps, the outer bundle detaches from the central bundle without significant slippage. Some snapshots of the process are shown in figure **3**.

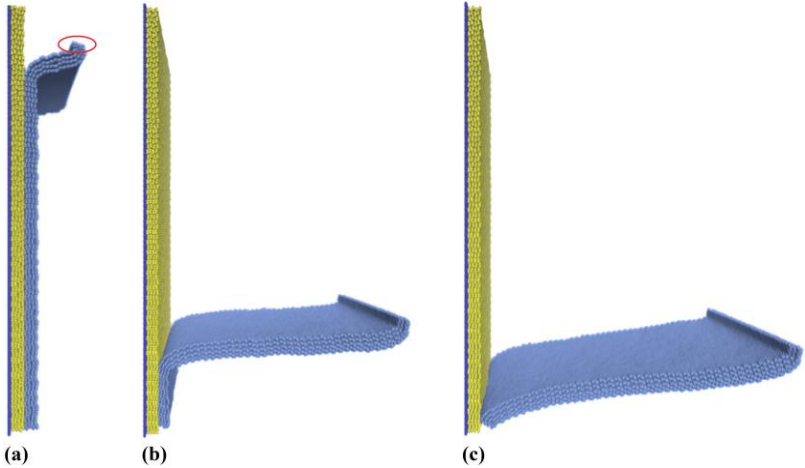

**Figure 3.** Lateral views of snapshots of the simulation of pulling out the outer bundle (blue) with 0.01 Å/ps in contact with the central bundle (yellow). Snapshots of the system **(a)** few moments after the beginning, **(b)** close to the end, and **(c)** at the final stage of the simulation of the process. Inside red circle, the region of atoms of the outer bundle to which the pulling out velocity is applied.

Figure **3** shows snapshots of the simulated FS pulling out process in the original configuration, i. e., the one shown in figure **2**(b). We can clearly see that the first (outer) bundle simply detached from the central bundle and, consequently, from the CNT forest, so not forming the CNTF as predicted by M1 model. This confirms the work of Machado, Legoas and Galvão [28] and proves that the simple M1 model is not able to explain the formation of CNTFs from the FS method. This result is not surprising because if we analyse the forces on the central bundle during the pulling out of the outer bundle, although this one attracts the central bundle by vdW forces, the graphene plane, representing the bundles located on the left side of the central bundle, also attracts the central bundle. In order to start detaching from the rest of the forest, the central bundle should bend at its extremity towards the pulling out direction as happened to the outer bundle at the beginning of the process (figure **3**(a)). There is an energy cost to bend the bundle and the interplay between equally intense vdW forces on both sides of the central bundle is not capable to make it being pulled out.

Since the conditions for the central bundle to be pulled out involves the forces on both sides of it, we performed an additional test using the configuration depicted in figure **2**(c). In this configuration, a bottom portion of the graphene plane representing the CNT bundle just on the left side of the central bundle, is cut to simulate a situation in which, at least close to the extremity of the bundle, there will be only attractive forces from the outer bundle. The question to be verified is that if it will be enough to have forces only on one side of the central bundle in order to pull it out towards the formation of continuous CNTFs like shown in figure **1**. The results for the quickest pulling out velocity are shown in figure **4**. In order to gain time, this simulation started from a frame where the outer bundle is connected to the central bundle close to the cut part of the graphene plane (figure **4**(a)).

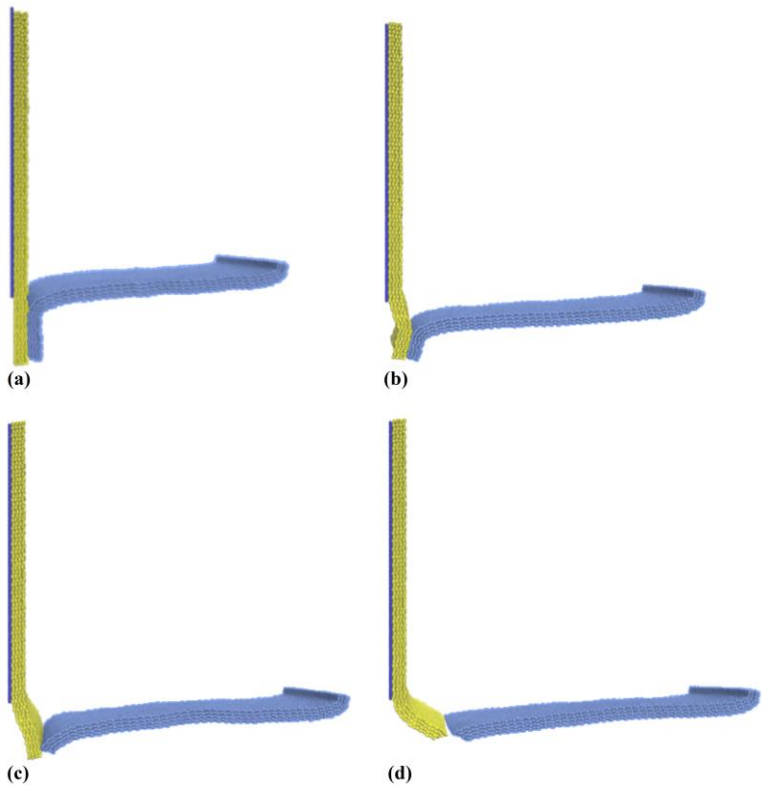

**Figure 4.** Lateral views of snapshots of the simulation of pulling out the outer bundle (blue) with 1 Å/ps in contact with the central bundle (yellow) in the configuration shown in figure **2**(c). **(a)** Outer bundle close to the beginning of the cut part of the graphene plane, **(b)** an intermediary frame where the central bundle started to bend, and **(c)** and **(d)** advanced stages of the simulation where it is possible to see that the contact area between outer and central bundles decreased and the outer bundle detached from the latter.

Figure **4** shows that the part of the outer bundle that interacts via vdW forces with the central bundle close to its extremity, now, has enough force to bend the central bundle. However, as the figures **4**(c) and **4**(d) show, the outer bundle ended up detached from the central bundle. Their contact area decreased during the process. The results can be summarized in two: i) it is possible for the bundle being pulled out to bend the central or next bundle as long as the extremity of the latter is not bonded to the rest of the CNT forest; ii) the M1 model is really not capable to describe the FS method. One of conclusions of Machado, Legoas and Galvão [28] is that the density of CNT forest should play an important role in its level of spinnability. The above results agree with that.

In spite of the lack of success for the formation of a CNTF from our second configuration of the M1 model, we believe that the size of the cut part of the graphene plane on the left side of the central bundle might be an important factor since it defines the radius of curvature that the central bundle might hold in order to be pulled out by vdW forces from the outer bundle. We believe that larger this space, larger the amount of contact area between the bundles at the extremity of the CNT forests, what might help promote the process. A systematically study of the FS process as a function of this cut size will be subject of a future publication.

## CONCLUSION

In this work, we presented MD simulations of one of the models for the FS dry-draw process of producing CNTFs. We showed that the simplest FS method model is not capable to describe the pulling out phenomenon. We also tested a different configuration with a cut about the end of the left side of the CNT bundle that is expected to be pulled out by the one being pulled, and found out that although the former gets bent by the latter, it is still not enough to form a CNTF like shown in the right part of figure **1**. Additional investigation of different sizes of this last configuration as well as definition of different approaches to simulate the models M2 and M3 are going on and will be subject of future publications.

## ACKNOWLEDGMENTS

AFF is a fellow of the Brazilian Agency CNPq-Brazil (303284/2021-8) and acknowledges grants #2020/02044-9 from São Paulo Research Foundation (FAPESP) and #2543/22 from FAEPEX/UNICAMP. This work used resources of the "Centro Nacional de Processamento de Alto Desempenho em São Paulo (CENAPAD-SP)" and of the John David Rogers Computing Center (CCJDR) in the "Gleb Wataghin" Institute of Physics, University of Campinas.

## CONFLICT OF INTEREST STATEMENT

On behalf of all authors, the corresponding author states that there is no conflict of interest.

## DATA AVAILABILITY STATEMENT

Data available on request from the authors.

## REFERENCES

[1] S. Iijima, *Nature* **354**, 56 (1991). https://doi.org/10.1038/354056a0
[2] F. Yang, M. Wang, D. Zhang, J. Yang, M. Zheng and Y. Li, *Chem. Rev.* **120**, 2693 (2020). https://doi.org/10.1021/acs.chemrev.9b00835
[3] B. Peng, M. Locascio, P. Zapol, S. Li, S. L. Mielke, G. C. Schatz and H. D. Espinosa, *Nat. Nanotechnology* **3**, 626 (2008). https://doi.org/10.1038/nnano.2008.211
[4] B. Q. Wei, R. Vajtai and P. M. Ajayan, *Appl. Phys. Lett.* **79**, 1172 (2001). https://doi.org/10.1063/1.1396632
[5] E. Pop, D. Mann, Q. Wang, K. Goodson and H. Dai, *Nano Lett.* **6**, 96 (2006). https://doi.org/10.1021/nl052145f
[6] M. F. L. De Volder, S. H. Tawfick, R. H. Baughman and A. J. Hart, *Science* **339**, 535 (2013). https://doi.org/10.1126/science.1222453
[7] K. Jensen, J. Weldon, H. Garcia and A. Zettl, *Nano Lett*. **7**, 3508 (2007). https://doi.org/10.1021/nl0721113

[8] M. M. Shulaker, G. Hills, N. Patil, H. Wei, H. -Y. Chen, H. -S. P. Wong and S. Mitra, *Nature* **501**, 526 (2013). https://doi.org/10.1038/nature12502
[9] X. Ge, M. Fu, X Niu and X. Kong, *Ceramics International* **46**, 26557 (2020). https://doi.org/10.1016/j.ceramint.2020.07.123
[10] R. Chadar, O. Afzal, S. M. Alqahtani and P. Kesharwani, *Colloids and Surfaces B: Biointerfaces* **208**, 112044 (2021). https://doi.org/10.1016/j.colsurfb.2021.112044
[11] T. E. Cantuario and A. F. Fonseca, *Annalen Der Physik* **531**, 1800502 (2019). https://doi.org/10.1002/andp.201800502
[12] T. N. Y. Silva and A. F. Fonseca, *Phys. Rev. B* **106**, 165413 (2022). https://doi.org/10.1103/PhysRevB.106.165413
[13] A. D. Avery, B. H. Zhou, J. Lee, E. -S. Lee, E. M. Miller, R. Ihly, D. Wesenberg, K. S. Mistry, S. L. Guillot, B. L. Zink, Y.-H. Kim, J. L. Blackburn and A. J. Ferguson, *Nat. Energy* **1**, 16033 (2016). https://doi.org/10.1038/nenergy.2016.33
[14] J. L. Blackburn, *ACS Energy Lett.* **2**, 1598 (2017). https://doi.org/10.1021/acsenergylett.7b00228
[15] I. A. Kinloch, J. Suhr, J. Lou, R. J. Young and P. M. Ajayan, *Science* **362**, 547 (2018). https://doi.org/10.1126/science.aat7439
[16] R. J. Headrick, D. E. Tsentalovich, J. Berdegué, E. A. Bengio, L. Liberman, O. Kleinerman, M. S. Lucas, Y. Talmon and M. Pasquali, *Adv. Mater.* **30**, 1704482 (2018). https://doi.org/10.1002/adma.201704482
[17] P. Dariyal, A. K. Arya, B. P. Singh and S. R. Dhakate, *J. Mater. Sci.* **56**, 1087 (2021). https://doi.org/10.1007/s10853-020-05304-z
[18] B. Natarajan, *Composites Science and Technology* **225**, 109501 (2022). https://doi.org/10.1016/j.compscitech.2022.109501
[19] F. Wang, S. Zhao, Q. Jiang, R. Li, Y. Zhao, Y. Huang, X. Wu, B. Wang and R. Zhang, *Cell Reports Physical Science* **3**, 100989 (2022). https://doi.org/10.1016/j.xcrp.2022.100989
[20] K. Jiang, Q. Li and S. Fan, *Nature* **419**, 801 (2002). https://doi.org/10.1038/419801a
[21] M. Zhang, K. R. Atkinson and R. H. Baughman, *Science* **306**, 1358 (2004). https://doi.org/10.1126/science.1104276
[22] X. Zhang, K. Jiang, C. Feng, P. Liu, L. Zhang, J. Kong, T. Zhang, Q. Li and S. Fan, *Adv. Mater.* **18**, 1505 (2006). https://doi.org/10.1002/adma.200502528
[23] Y. Nakayama, *Jpn. J. Appl. Phys.* **47**, 8149 (2008). https://doi.org/10.1143/JJAP.47.8149
[24] A. A. Kuznetsov, A. F. Fonseca, R. H. Baughman and A. A. Zakhidov, *ACS Nano* **5**, 985 (2011). https://doi.org/10.1021/nn102405u
[25] C. Zhu, C. Cheng, Y. H. He, L. Wang, T. L. Wong, K. K. Fung and N. Wang, *Carbon* **49**, 4996 (2011). https://doi.org/10.1016/j.carbon.2011.07.014
[26] D. W. Brenner, O. A. Shenderova, J. A. Harrison, S. J. Stuart, B. Ni and S. B. Sinnott, *J. Phys.: Condens. Matter* **14**, 783 (2002). https://doi.org/10.1088/0953-8984/14/4/312
[27] A. P. Thompson, H. M. Aktulga, R. Berger, D. S. Bolintineanu, W. M Brown, P. S. Crozier, P. J. in 't Veld, A. Kohlmeyer, S. G. Moore, T. D. Nguyen, R. Shan, M. J. Stevens, J. Tranchida, C. Trott, and S. J. Plimpton, *Comput. Phys. Commun.* **271**, 108171 (2022). https://doi.org/10.1016/j.cpc.2021.108171
[28] L. D. Machado, S. B. Legoas and D. S. Galvão, *MRS Online Proceedings Library* **1407**, 710 (2012). https://doi.org/10.1557/opl.2012.710
[29] R. R. Del Grande, A. F. Fonseca and R. B. Capaz, *Carbon* **159**, 161 (2020). https://doi.org/10.1016/j.carbon.2019.12.030